\documentclass[final,5p,times]{elsarticle}

\usepackage{amssymb}
\usepackage{amsmath}
\usepackage{graphicx}
\usepackage{lineno}
\usepackage{float}
\usepackage{afterpage}
\usepackage{placeins}
\usepackage{booktabs}
\usepackage{xcolor}
\usepackage[utf8]{inputenc}

\journal{Nuclear Instruments and Methods in Physics Research Section A}

\begin{document}

\begin{frontmatter}

\title{Implementation of Prototype Permanent Microundulators}

\author[label1]{E. Magory}
\author[label2]{V. L. Bratman}
\author[label2]{A. Steiner}
\author[label2]{N. Balal\corref{cor1}}
\ead{nezahb@ariel.ac.il}

\cortext[cor1]{Corresponding author}

\address[label1]{Department of Electrical and Electronic Engineering, Jerusalem College of Technology, Jerusalem, Israel}
\address[label2]{Department of Electrical and Electronic Engineering, and Schlesinger Family Center for Compact Accelerators, Ariel University, Ariel, Israel}

\begin{abstract}
We report on the implementation of two types of helical microundulators with a small period of 6 mm, containing axially magnetized helices fabricated from a single piece of rare-earth magnet. The first microundulator, formed by two oppositely magnetized longitudinal helices, produces a transverse on-axis field higher than 0.93 T, while the second, assembled from two oppositely longitudinally magnetized rare-earth helices and two pre-unmagnetized steel helices, produces a field of about 1.5 T. The apertures inside both microundulators are 1 mm, which required special field measurement techniques. When used in compact FELs in the terahertz to X-ray range, such microundulators can provide significantly higher electron oscillation velocity and radiated power than the previously proposed planar microundulators. An attractive example of an ultra-compact XFEL in SASE regime is presented, in which the use of the studied hybrid helical microundulator can provide a radiation power of 48 GW at a wavelength of about 5 Å.
\end{abstract}

\begin{keyword}
Free electron laser \sep Undulator \sep X-ray \sep Permanent magnet \sep Helical undulator
\end{keyword}

\end{frontmatter}

\section{Introduction}

At present, Free Electron Lasers (FELs) successfully compete with conventional quantum lasers in a number of frequency ranges and are ahead of them in mastering short waves up to hard X-rays. XFELs \cite{ref1,ref2,ref3,ref4,ref5,ref6} are unique sources of powerful coherent X-ray radiation, which currently provide for the rapid development of many important areas of science. However, all existing such sources are large and expensive installations, which include accelerators of electron beams with very high energies of 5--18 GeV, available to a relatively small range of researchers. In order to expand the scope of research, many papers propose more compact XFELs with lower electron energies and dimensions. To implement such sources, highly non-trivial methods have been proposed for the formation and acceleration of short high-density electron bunches with a high energy gain rate, as well as with very small particle energy spreads and emittances. In most of these projects, high-field microundulators also play a key role. For example, to design ultra-compact soft and hard UC-XFELs with a total length of less than 50 m, a new version of permanent Halbach \cite{ref7} planar microundulators with a strong on-axis field was proposed to achieve radiation wavelengths of 10 and 1.6 Å at relatively low electron energies \cite{ref8}. As an alternative, our papers \cite{ref9,ref10,ref11,ref12} propose permanent helical microundulators based on the use of magnetized rare-earth helices, which allow for significantly higher power and efficiency of XFELs compared to planar ones. Section 2 of this paper presents the results of calculations, manufacturing, and experimental study of two types of such microundulators with a period of 6 mm and a high field on the axis up to 1.5 T. With the same field strength and a shorter period, this simply implemented prototype of an undulator, assembled from helices and working at room temperature, is easier to manufacture and operate than modern cryogenic permanent magnet undulators \cite{ref13,ref14,ref15,ref16,ref17}. Section 3 numerically studies promising examples of compact XFELs with planar and hybrid helical microundulators, as well as with electron bunches, the production of which can be based on the ideology and design developed in \cite{ref8}. Section 4 summarizes the results of this work.

\section{Periodic Systems from a Single Piece of Magnet}

\subsection{Magnetized combs and helices}

One of the most important new elements in promising projects of ultra-compact X-ray FELs \cite{ref8,ref18} is the permanent planar Halbach-type \cite{ref7} microundulator proposed in \cite{ref8,ref19}. It is assumed that it will be assembled from original magnetized combs, each of which is made not from many small magnets, but from a single piece of rare-earth magnet, which should reduce the non-uniformity of magnetization and simplify the assembly of undulators. As an alternative, our works \cite{ref9,ref10,ref11,ref12} propose helical microundulators -- rare-earth helices, which are also made of a solid magnet, but provide for FELs stronger average fields and, therefore, significantly higher amplitudes of electron oscillations and radiation power. For the precision manufacture of helices from brittle rare-earth materials, we use Wire Electrical Discharge Machining (WEDM) \cite{ref11}, also discussed as a possible method for manufacturing planar combs in \cite{ref8}.

\begin{figure*}[!t]
\centering
\includegraphics[width=0.85\textwidth]{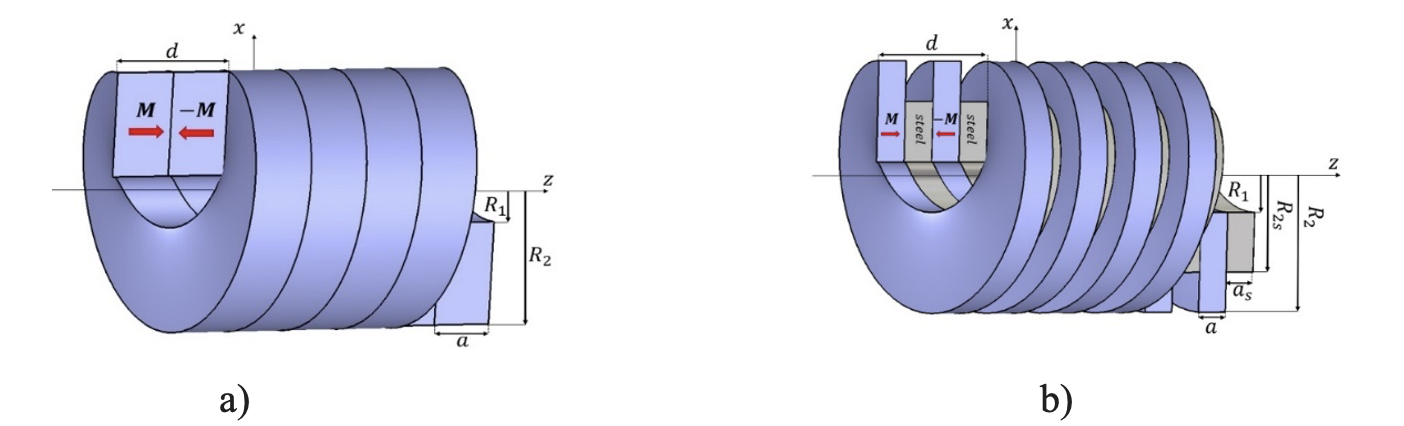}
\caption{Helical undulators, assembled from a) two identical, oppositely longitudinally magnetized helices, shifted by half the period, b) two identical, oppositely longitudinally pre-magnetized helices alternating with two pre-unmagnetized steel helices (hybrid helical undulator).}
\label{fig1}
\end{figure*}

A rare-earth helix uniformly magnetized in the longitudinal or radial directions (along its axis and perpendicular to it, respectively) creates a helical magnetic field on the axis with a period equal to the helix period $d$. Two identical but oppositely magnetized helices, shifted relative to each other along the axis by half the period, create a transverse field twice as large on the axis and form a simple undulator (Fig.~\ref{fig1}a).

The "ideal" Halbach-type undulator, consisting of alternating two longitudinally and two radially oppositely magnetized helices with widths $d$/4 each \cite{ref10,ref12}, which is a direct generalization of the planar Halbach undulator \cite{ref7} to the helical one, creates a stronger internal and almost zero external field. At small inner radii, radially magnetized helices create a significantly higher field than longitudinally magnetized ones. However, given the difficulties of implementing uniformly radially magnetized helices, it is much easier to replace them with pre-unmagnetized helices made of steel or vanadium permendur (Fig.~\ref{fig1}b), which redistribute the magnetic flux from the longitudinally pre-magnetized rare-earth helices in the radial direction \cite{ref12}. It should be noted that in \cite{ref20} a hybrid helical permanent high-field undulator was proposed, consisting of individual helical vanadium permendur poles sandwiched between helical NeFeB blocks, capable of generating adjustable fields.

In this Section, we report on the fabrication, calculation, magnetization, assembling, and measurement of the parameters for two mentioned types of helical microundulators (Fig.~\ref{fig1}) with periods $d$=6 mm and inner hole radius $R_1$=0.5 mm, consisting of 1) two oppositely longitudinally magnetized NdFeB helices with an equal longitudinal width of 0.5 $d$ and 2) two oppositely longitudinally pre-magnetized NdFeB helices alternating with two pre-unmagnetized high-permeability helices made of steel (hybrid helical microundulator).

\begin{figure*}[!t]
\centering
\includegraphics[width=0.85\textwidth]{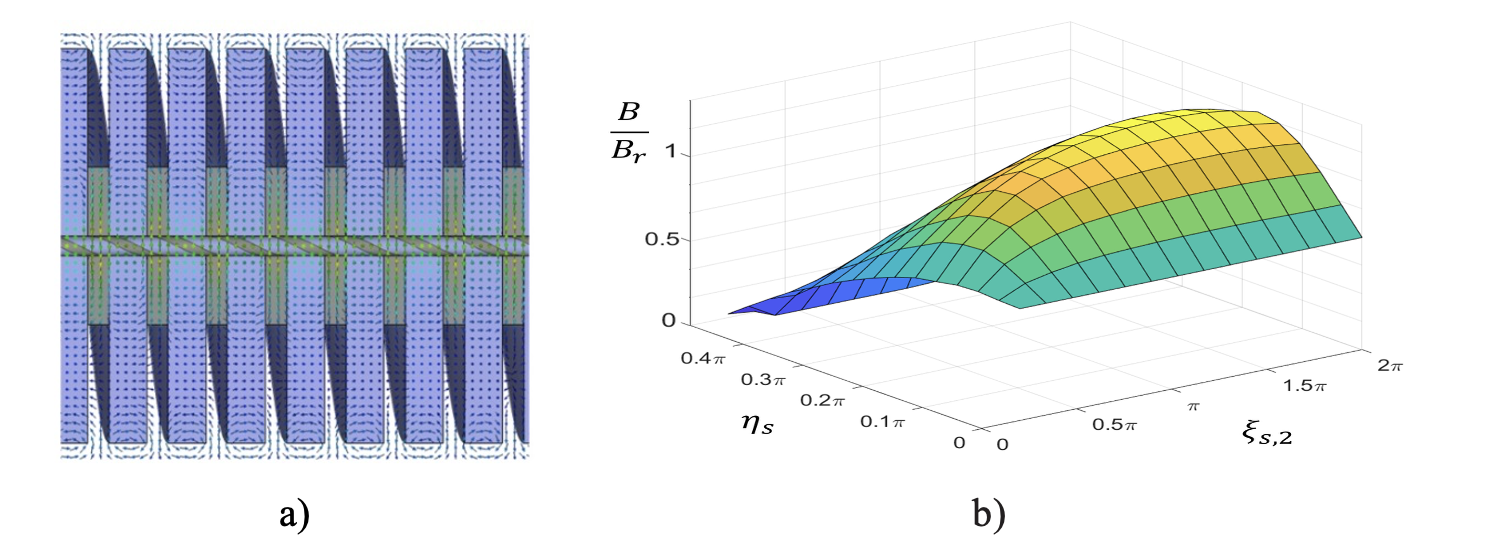}
\caption{Modeling of a hybrid helical undulator using the CST Microwave Studio \cite{ref21}: a) steel helical inserts redirect the magnetic flux radially, forming a field structure near the axis close to the field in an "ideal" Halbach-type helical undulator, b) optimization of the magnetic field on the axis using two dimensionless parameters of the steel helices - the outer radius $\xi_{s2} = \frac{2\pi R_s}{d}$ and width $\eta_s = \frac{\pi a_s}{d}$.}
\label{fig2}
\end{figure*}

By redirecting and concentrating the magnetic flux generated by longitudinally pre-magnetized rare-earth helices, the steel helices provide a field structure inside such a hybrid helical undulator (Fig.~\ref{fig2}a) that is very close to the structure of an "ideal" helical undulator \cite{ref10,ref11,ref12}.

\subsection{Fabrication and magnetization of helices}

\begin{figure*}[!t]
\centering
\includegraphics[width=0.85\textwidth]{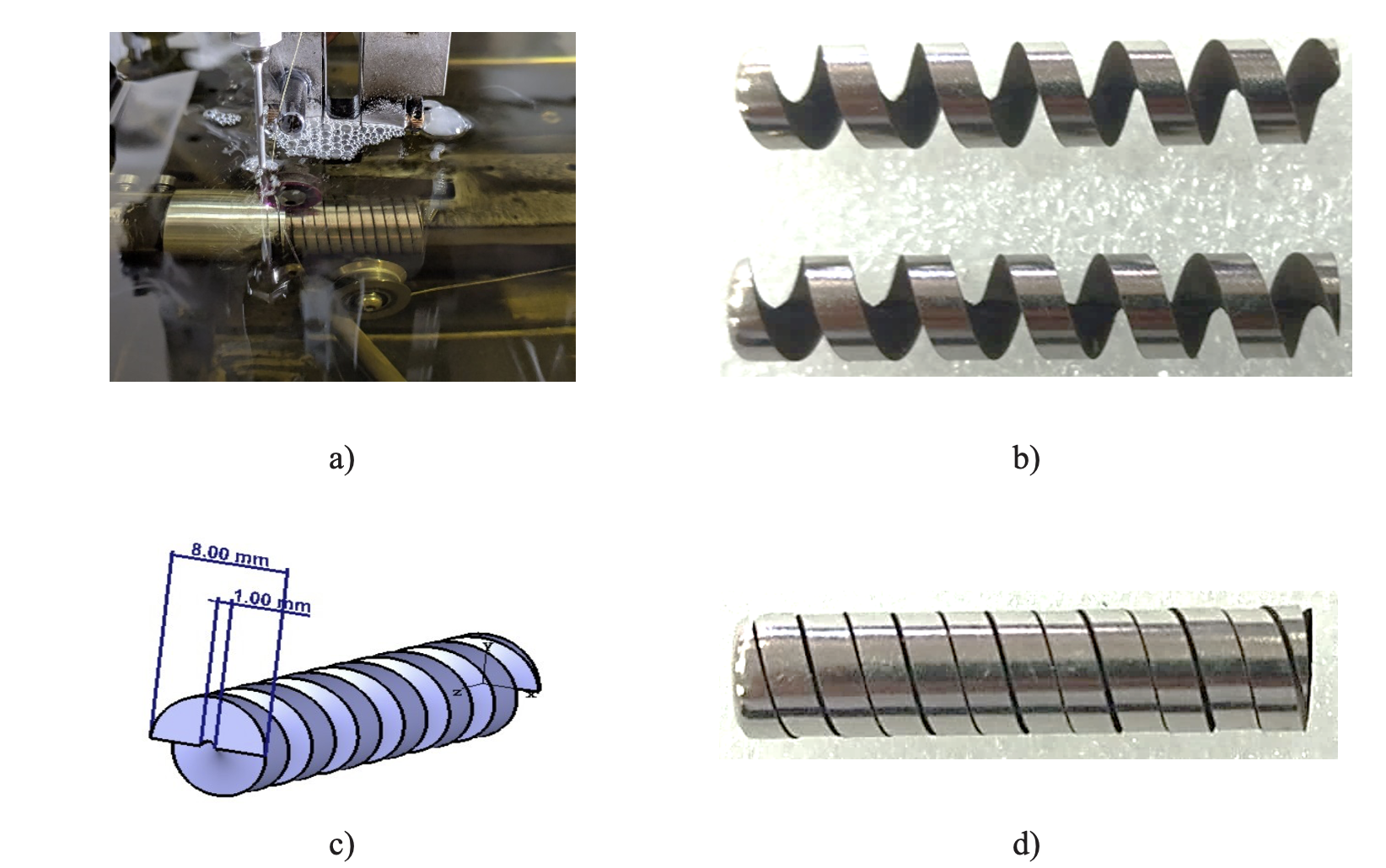}
\caption{a) WEDM cutting of a helical groove in an NdFeB pipe provides two identical helices b) and c) with a period of 6 mm and a length of 50 mm each, d) the final microundulator assembled from two oppositely longitudinally magnetized helices.}
\label{fig3}
\end{figure*}

Using Wire EDM and EDM drilling, we manufacture rectangular small-period rare-earth and steel helices from solid cylinders (Fig.~\ref{fig3}). First, a 1-mm diameter axial bore is drilled in a solid NdFeB or steel cylinder by EDM using a brass electrode with a diameter of about 1 mm. EDM was preferred over mechanical drilling due to the brittleness of the NdFeB magnets, which are subject to cracking. The EDM process removes material by localized evaporation, which ensures precise bore formation with minimum mechanical stress and thermal damage. WEDM was then used to fabricate both helices from the pipe (Fig.~\ref{fig3}) using a thin wire brass electrode with a diameter of 0.25 mm, which follows a programmed path and destroys the material with rapid electrical discharges. This technology is particularly suited for hard and brittle magnetic materials, providing micron-level dimensional accuracy. By coordinating the rotation of the magnet cylinder with the linear translation of the wire, a continuous helical groove was cut along the entire length of the magnet (Fig.~\ref{fig3}).

The cutting program was designed to create a double-helix pattern with wire wound on a cylinder with a pitch of 6 mm, which effectively divided the magnet into two intertwined helical sections. WEDM was carried out with the workpiece immersed in dielectric oil to improve process stability and cooling. To minimize surface roughness and prevent the formation of microcracks in the NdFeB material, the following optimal combination of cutting parameters was experimentally selected: pulse current 12 A, pulse duration $10$--$40~\mu$s, wire tension 1.5 N, and feed rate 1.5 m/min.

The manufactured NdFeB helices are magnetized to saturation in the longitudinal direction (along the axis) in the field of a pulsed solenoid with a field of about 2-3 T and a pulse duration of several milliseconds.

\subsection{Microundulator assembled from two NdFeB helices with opposite longitudinal magnetization}

To fabricate a simple undulator with a strong field, two identical and oppositely longitudinally magnetized helices, each having a width equal to half the period, $a = d/2$, are used. The field of an infinite helix on the axis is \cite{ref22}

\begin{equation}
\mathbf{B} = B_L(\hat{x}\sin\zeta + \hat{y}\sin\zeta), \quad B_L = B_r\frac{1}{\pi}\int_{\xi_1}^{\xi_2}\xi K_1(\xi)d\xi.
\label{eq1}
\end{equation}

Here, $\zeta = hz$, $h$ = 2$\pi$/$d$, $d$ is the period of the helix and undulator, $B_r$ is the remnant magnetization of the rare-earth magnet, $\xi_{1,2} = \frac{2\pi R_{1,2}}{d}$, $R_{1,2}$ are the inner and outer radii of the helices, and $K_1(\xi)$ is a first-order MacDonald function. Note that Eq.~(\ref{eq1}) gives a satisfactory estimate of the helical field even for a small number of undulator periods down to $N\sim 5$. The helical field on the $z$ axis of the infinite length system from two helices is obviously twice as large as the field created by a single helix.

To fabricate two identical helices with a period of 6 mm and the same longitudinal thickness $a = 3$ mm, a pipe with inner and outer radii $R_1 = 0.5$ mm and $R_2 = 4$ mm, respectively, and a length of 50 mm, fabricated from a NdFeB cylinder with a high nominal remnant magnetization in the range of $B_r$=1.43-1.48 T, was used. The microundulator of two oppositely magnetized helices (Fig.~\ref{fig3}d) was assembled by screwing one of the helices into the other. According to Eq.~(\ref{eq1}), the helical field on the axis of such a microundulator of infinite length is 0.95-0.98 T. CST calculations for the system with a length of 50 mm give 0.93-0.95 T.

\begin{figure*}[!t]
\centering
\includegraphics[width=0.85\textwidth]{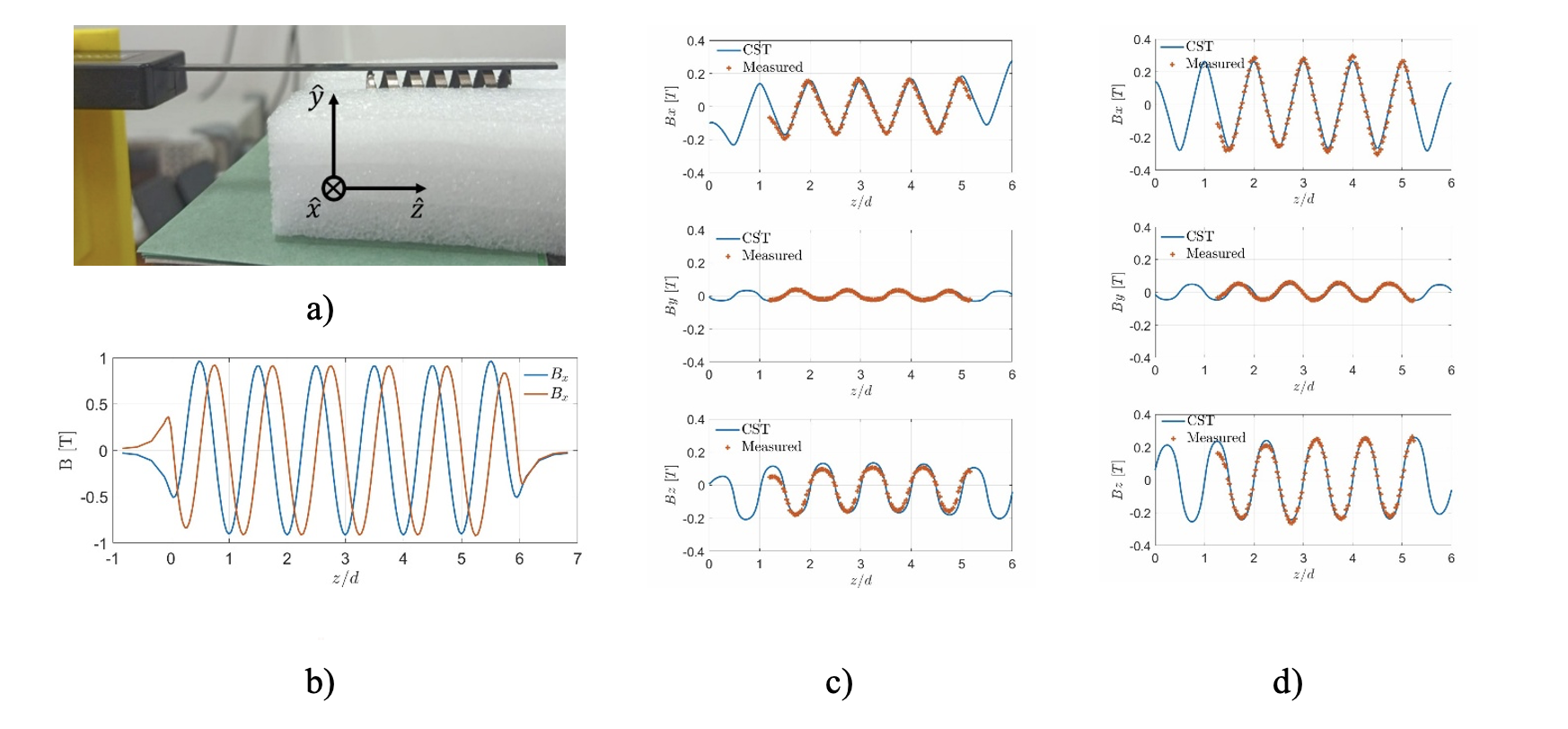}
\caption{a) Measuring the field outside the helices: the Hall sensor is located at the distance $r$ = 5 mm from the helix axis, b) results of CST calculations for the field on the axis, c) and d) comparison of the results of CST calculations with the measured magnetic field for a single helix and two oppositely magnetized helices at the distance $r$ = 5 mm from the axis.}
\label{fig4}
\end{figure*}

Because of the small diameter of the inner hole, direct measurement of the field on the axis of the helices is complex and requires a very small magnetic sensor. Instead, we used the fact that a single helix creates a strong outer field and performed measurements from the outside at a minimum distance from the outer surface of the helix, which corresponded to the distance $r$ = 5 mm from the axis (Fig.~\ref{fig4}). In this case, the Hall sensor inside the Senis 3MTS USB handheld Tesla meter was carefully aligned with the edge of the helix. The measurement results for all components of the magnetic field of the helix are in good agreement with the calculations (Fig.~\ref{fig4}c).

Unlike the Halbach array and similar to the single helix, the double-helix microundulator produces a significant external field. The measured field outside the microundulator also agrees well with the CST simulation results (Fig.~\ref{fig4}d). These measurements and comparisons show that the magnitude of the helical field on the axis exceeds 0.93 T.

The next section will also describe a more complex but direct method of measuring the field closer to the axis by placing the axially moving probe perpendicular to the axis inside a single synchronously rotating magnetized helix.

\subsection{Hybrid helical microundulator}

As was already mentioned, the field of the hybrid helical undulator on its axis is close to the field of the "ideal" infinite helical Halbach-type undulator, which in turn is described by the formulas \cite{ref10}

\begin{equation}
\mathbf{B} = B_H\left( \hat{x}\sin\zeta + \hat{y}\sin\zeta \right), \quad B_H = B_r\frac{\sqrt{2}}{\pi}\int_{\xi_1}^{\xi_2}\xi\left[ K_1(\xi) - K_1'(\xi) \right]d\xi,
\label{eq2}
\end{equation}

where $K_1'(\xi)$ is the derivative of a first-order MacDonald function.

According to CST calculations, a hybrid undulator with equal longitudinal thicknesses of rare-earth and steel helices $a = a_s = 0.25d$ and optimized outer radius of steel helices R$_{2s}$ (Fig.~\ref{fig2}b) provides a transverse field on the axis that is only 7\% smaller than that of the "ideal" helical Halbach undulator. Optimization of the thicknesses (Fig.~\ref{fig2}b) permits one to obtain the same value of the on-axis field as in an infinite "ideal" Halbach-type undulator; the corresponding thicknesses of the helices are $a$=0.32 $d$ and $a_s$=0.18 $d$. It is even more important that the corresponding increase in the width of rare-earth helices provides a more robust design of the microundulator. The optimized parameters of a steel helix are as follows: the dimensionless outer radius is $\xi_{s2} = 1.3\pi$ and the longitudinal thickness is $\eta_s = 0.18$ versus $\eta = 0.32 d$ for the NdFeB component.

\begin{figure*}[!t]
\centering
\includegraphics[width=0.85\textwidth]{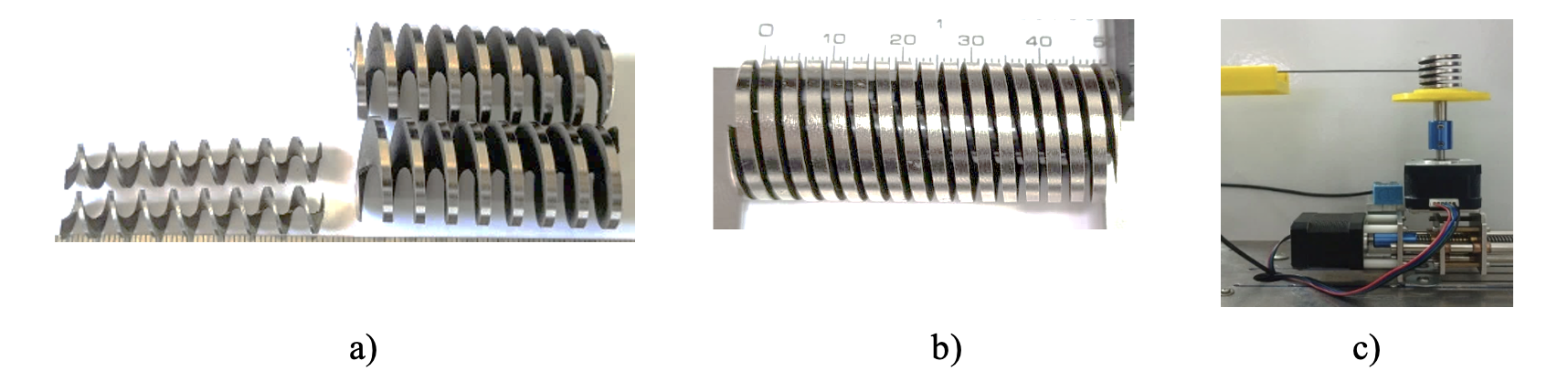}
\caption{a) Steel and NdFeB helices, b) assembled hybrid helical microundulator, c) measuring the inner field of a single magnetized helix with a small axial hole: the probe immersed inside the rotating helix moves along its axis synchronously with its rotation.}
\label{fig5}
\end{figure*}

To fabricate a prototype of the optimized hybrid microundulator with a period of 6 mm and an axial field of about 1.5 T, the helices were made of solid NdFeB and steel cylinders with outer radii $R_2$ = 10 mm and $R_{2s}$ = 4 mm, respectively, and length $L$ = 50 mm. The outer radius of the NdFeB cylinder is fairly large to compensate for a relatively low value of its remnant magnetization $B_r$=1.2 T. Holes of 1 mm diameter were drilled along the axis of both cylinders. Then the cylinders with holes were cut into two helices, each with longitudinal thicknesses of 1.92 mm and 1.08 mm, using WEDM (Figs.~\ref{fig3}a and~\ref{fig5}). Cutting was performed in oil to improve machining stability. Aluminum spacers were fabricated to attach the components to the rotor. The optimized cutting modes were defined for each material. For the NdFeB, a current of 12 A was used with a duty cycle of 10 to 40 µs for a wire tension of 1.5 N and a feed rate 1.5 m/min. The longitudinal cutting speed was approximately 80 µm/min. Both materials were cut using a 0.25-mm brass wire.

\begin{figure*}[!t]
\centering
\includegraphics[width=0.85\textwidth]{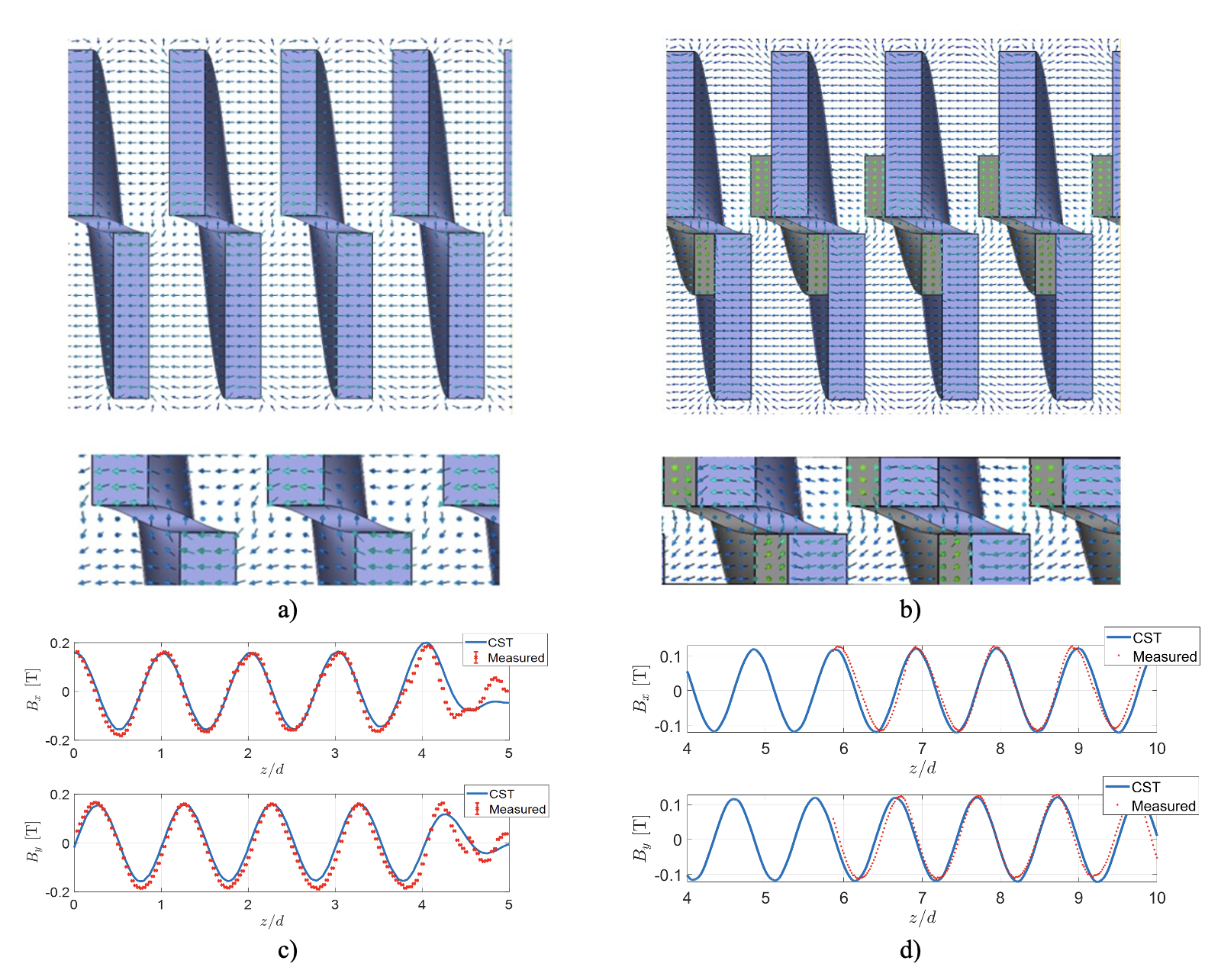}
\caption{Comparison of calculations and measurements of the transverse field: a) and b) field distributions for a single longitudinally pre-magnetized NdFeB helix and for the same helix with a pre-unmagnetized steel helix (top) and an enlarged image of the region near the axis (bottom); c) and d) field components at distances of 1 and 3.5 mm from the axis for a single pre-magnetized NdFeB helix and for the same helix with a pre-unmagnetized steel helix, respectively.}
\label{fig6}
\end{figure*}

The obtained two rare-earth and two steel helices (Fig.~\ref{fig5}a) were used to assemble a prototype hybrid microundulator (Fig.~\ref{fig5}b). The NdFeB helices were previously magnetized, taking precautions to avoid mechanical damage. After assembling the NdFeB helices alternating with steel ones, a strong helical field is created inside the hole, but the outer fields outside the microundulator are compensated. That is why the field of a single pre-magnetized NdFeB helix, as well as a system consisting of a single pre-magnetized NdFeB helix and a single pre-unmagnetized steel helix, were measured in the experiment and compared with CST calculations (Fig.~\ref{fig6}). For this purpose, we used a measurement scheme differing from that described for the microundulator assembled from two NdFeB helices. Namely, the probe of a compact 3D Hall sensor (Teslameter 3MTS Senis) was immersed perpendicular to the axis into the measured helix from its outer side and the sensor moved along the rotating helix (Fig.~\ref{fig5}c). The probe moved along the $z$ axis, while the helix rotated at an angle $\Delta\theta = \frac{2\pi}{d} \cdot \Delta z$. Such synchronization allowed the field to be sampled near the helix axis. The measured field of a single helix agrees very well with the calculated one (Fig.~\ref{fig6}). Discrepancies of the order of several percent probably arise due to a slight non-uniformity of magnetization and inaccuracy of electrical discharge cutting, but a particularly important reason is the difficulty of accurately determining the location and direction of the measuring system inside the helix.

The final assembled eight-period hybrid helical microundulator (Fig.~\ref{fig5}b) consists of two oppositely longitudinally pre-magnetized NdFeB helices alternating with two pre-unmagnetized steel helices. A central 1-mm diameter hole runs through the entire device, providing space for the electron beam and radiation in the FELs. The result of the measurement and calculation allows us to conclude that the field on the axis of the hybrid microundulator is 1.5 T with an accuracy of about 5\%.

\section{Efficiency of ultra-compact X-ray FELs with permanent planar and helical strong-field microundulators}

Let us demonstrate the capabilities of the studied microundulator for the implementation of an X-ray FEL, using the ideology developed in \cite{ref8}. The ultra-compact XFELs with relatively modest electron energies of 1 and 1.6 GeV and small undulator periods of 6.5 and 3 mm, respectively, were designed in \cite{ref8} to produce powerful radiation in the soft and hard X-ray ranges with wavelengths of 10 and 1.6 Å. Our calculations \cite{ref12} based on the steady-state version of the Genesis code \cite{ref23,ref24} demonstrate that replacing planar undulators with our proposed helical undulators with the same period and a field strength of 1 T in each of both transverse Cartesian components can lead to a significant increase in the output radiation power. In this Section, we will show that increasing the field strength to 1.5 T, obtained in our experiment with a 6-mm period hybrid microundulator (Section 2), allows us to further increase the radiation power, bringing it to values planned for large installations.

\begin{table}[!htbp]
\centering
\caption{Some design parameters of a hard UC-XFEL with a planar microundulator \cite{ref8}}
\label{tab1}
\begin{tabular}{ll}
\toprule
Electron energy, GeV & 1.6 \\
Energy spread, \% & 0.03 \\
Microbunch charge, pC & 4.7 \\
Microbunch r.m.s. length, nm & 140 \\
Peak current, kA & 4.0 \\
Peak microbunch power, TW & 6.4 \\
Mean spot size $\sigma_r$ & 4.1 \\
Undulator period $d$, mm & 3 \\
Peak undulator field $B_P$ & 1 T \\
Undulator parameter $K$ & 0.28 \\
Radiation fundamental wavelength $\lambda_r$, Å & 1.6 \\
Pierce parameter $\rho$, 10$^{-3}$ & 0.78 \\
\bottomrule
\end{tabular}
\end{table}

The inner diameter of a 6-mm period microundulator with a strong field of 1.5 T, studied in Section 2, is 1 mm. Approximately the same gap value allows obtaining a field of 1 T with a period of 3 mm in the planar microundulator considered in \cite{ref8} for the hard UC-XFEL. At room temperatures and with the magnetic materials we used, a field of 1.5 T with a period of 3 mm can be obtained only with very small gaps of the order of 0.5 mm. Therefore, we have studied another attractive option - the radiation of a high-quality electron bunch with designed parameters of the hard UC-XFEL (Table~\ref{tab1}), but in planar Halbach and hybrid helical microundulators operating in the parameter range we have already mastered, namely, with a period of 6 mm and a strong peak field of 1.5 T (Table~\ref{tab2}).

\begin{table}[!htbp]
\centering
\caption{Modified parameters of the considered XFELs with planar and hybrid helical high-field microundulators}
\label{tab2}
\begin{tabular}{lll}
\toprule
 & Planar & Helical \\
\midrule
Undulator period $d$, mm & 6 & 6 \\
Peak undulator field, T & $B_P = 1.5$ & $B_H = 1.5$ \\
Undulator parameter $K$ & 0.59 & 0.84 \\
Pierce parameter $\rho$, 10$^{-3}$ & 2.8 & 4.1 \\
Radiation fundamental wavelength $\lambda_r$, Å & 4.1 & 5.3 \\
\bottomrule
\end{tabular}
\end{table}

In Tables~\ref{tab1} and~\ref{tab2} for planar and helical undulators,

\begin{equation}
K = \frac{eBd}{2\pi mc},
\label{eq3}
\end{equation}

where $e$ and $m$ are the electron charge and mass, $c$ is the speed of light, the field of a planar undulator on its axis is $\mathbf{B} = \hat{y}B_P\sin\zeta$, and $B = \frac{B_P}{\sqrt{2}}$ and $B = B_H$ are the r.m.s. values of the fields in planar and helical undulators, respectively,

\begin{equation}
\rho \approx \left[\frac{I}{4\pi^2 I_A} \cdot \frac{\gamma K^2}{\left(\sigma/\lambda^2\right)}\right]^{1/3}
\label{eq4}
\end{equation}

is the Pierce parameter in the simplest 1D theory (see, e.g., \cite{ref8,ref23,ref24}), $I$ is the electron current, $I_A$ is the Alfvén current, $\gamma$ is the Lorentz factor of the electrons, and

\begin{equation}
\lambda \approx \frac{d}{2\gamma^2}(1+K^2)
\label{eq5}
\end{equation}

is the radiation fundamental wavelength, i.e., the wavelength emitted by the electron bunch at the first undulator harmonic along the undulator axis in planar and helical undulators.

\begin{figure*}[!t]
\centering
\includegraphics[width=0.85\textwidth]{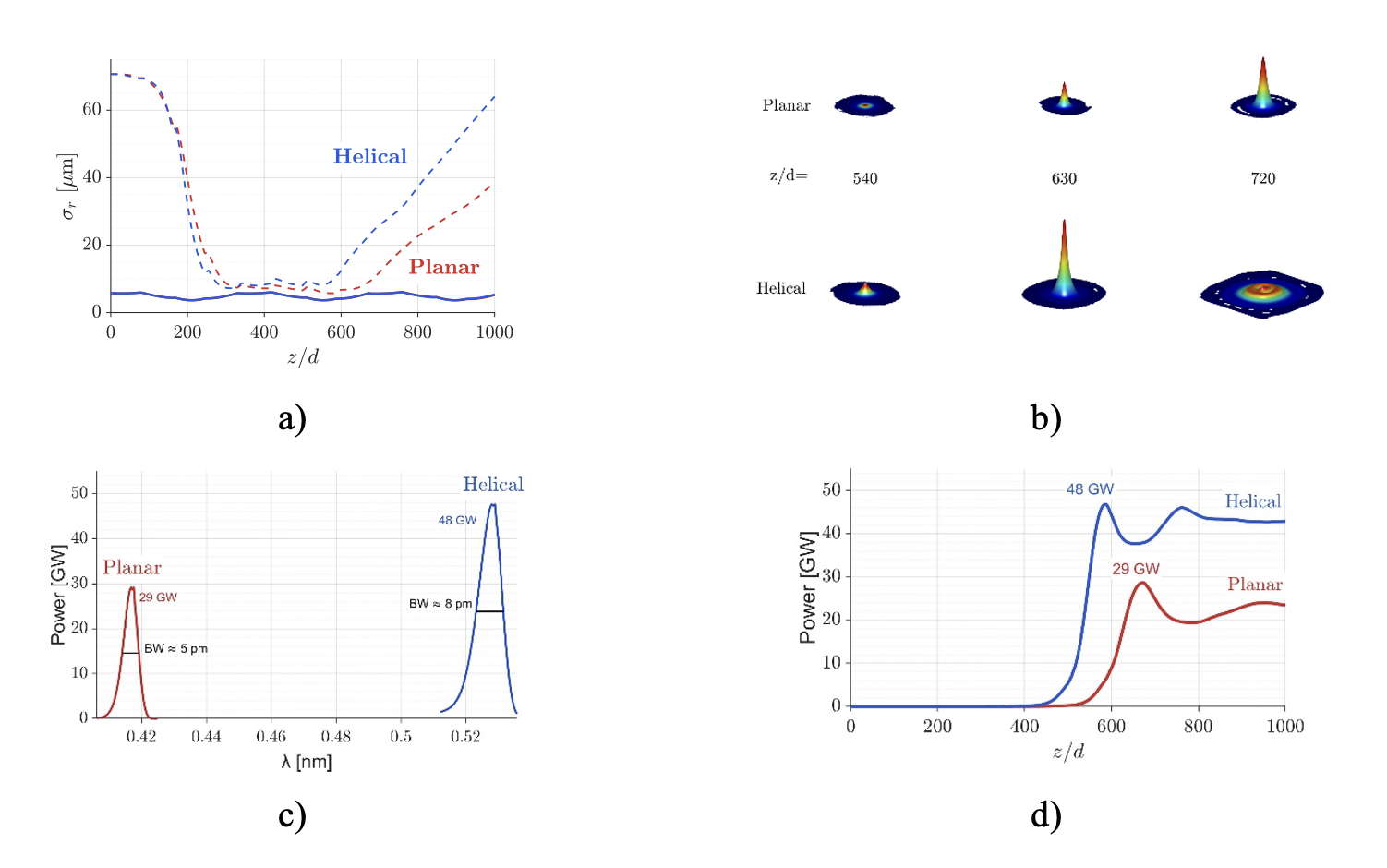}
\caption{a) Effective radial sizes of the focusing electron bunch and guiding X-ray radiation and b) evolution of the radiation intensity distribution along the interaction section of an XFEL, c) radiation spectra, and d) power for the electron bunch focused by a periodic quadrupole FODO array in helical and planar microundulators with field amplitudes of 1.5 T and a period of 6 mm.}
\label{fig7}
\end{figure*}

We numerically studied the operation of the XFEL at the first harmonic in the Self-Amplified Spontaneous Emission (SASE) regime using the steady-state version of the Genesis code \cite{ref23,ref24} (Fig.~\ref{fig7}). In this regime, the initial X-ray pulse is generated by the electron bunch at the stage of its spontaneous emission as a set of transverse modes. Then, as the electrons are microbunched under the action of this radiation, the resonant radiation with a decreasing number of modes and an increasing degree of transverse coherence is channeled in the optical guidance mode (Figs.~\ref{fig7}a and~\ref{fig7}b). In our calculations, it is assumed that in both compared planar and helical microundulators the electron bunch is focused by the same FODO system of strong-field magnetic quadrupoles, so that the r.m.s. transverse dimensions of the electron bunch are close to constant and are almost the same for these two types of microundulators (Fig.~\ref{fig7}a). When calculating the focusing system, we used the method described in \cite{ref25}.

In a strong field $B$=1.5 T, fairly large values of the undulator parameter $K$ and the Pierce parameter $\rho$ have been achieved (Table~\ref{tab2}). The Pierce parameter~(\ref{eq4}) gives an estimate of the order of magnitude of the radiation efficiency and the characteristic gain length $L_g $:

\begin{equation}
\eta = \frac{P}{P_b} \sim \rho, \quad L_g \approx \frac{d}{4\pi\sqrt{3}\rho},
\label{eq6}
\end{equation}

as well as the saturation length (see details and results of more complex 3D theory in, e.g., \cite{ref8,ref26,ref27}); here, $P$ and $P_b$ are the peak X-ray radiation and microbunch powers, respectively.

The calculated values of the resonance wavelengths of X-ray radiation 4.1 and 5.3 Å (Fig.~\ref{fig7}c) are close to the values determined by Eq.~(\ref{eq5}). The spectrum width corresponds to the gain saturation length. When the resonance condition is met for planar and helical undulators, the maximum X-ray radiation power in planar and helical microundulators reaches 29 and 48 GW (Fig.~\ref{fig7}d), which is close to or even exceeds the radiation power achieved to date in large installations with a significantly higher power of the electron microbunch. In accordance with the above simplest order-of-magnitude estimate~(\ref{eq6}), in the considered variants of ultra-compact XFELs with large Pierce parameters of 0.0028 and 0.004 for planar and helical microundulators, respectively, a high radiation efficiency of 0.45\% and 0.75\% can be achieved. Due to the large value of the undulator parameter $K$ in the considered variants, the Pierce parameter significantly exceeds the energy spread. Accordingly, with a significant increase in the spread from 0.03\% to 0.1\%, the calculated peak values of power and energy remain quite large: 20 GW, 0.3\% and 37 GW, 0.58\% for planar and helical microundulators, respectively.

\section{Conclusions}

In this paper, we develop the idea of a high-field microundulator assembled not from many miniature magnetic elements, but from several periodic magnetic structures, each made of a single piece of rare-earth magnetic material \cite{ref8,ref9,ref10,ref11,ref12,ref19}. The paper studies two prototype microundulators containing longitudinally (axially) magnetized NdFeB helices with a period of 6 mm and a through axial hole of 1 mm.

The simplest prototype microundulator, consisting of two helices with the same magnitude and opposite direction of magnetization, creates a transverse field of 0.93 T on the axis. Its disadvantage, as in the case of a planar "up-down" undulator, is the large value of the external field. However, measuring this field with sufficiently high accuracy and good agreement between the measurement results and the calculation makes it easy to find the field on the axis even with a small axial hole.

The prototype eight-period hybrid helical undulator assembled from two oppositely pre-magnetized NdFeB helices alternating with two pre-unmagnetized steel helices creates a field of 1.5 T on the axis and a virtually zero field outside. To find the field on the axis, measurements and comparisons with calculations were used for a single pre-magnetized NdFeB helix, as well as for the same helix and a single pre-unmagnetized steel helix, using a probe immersed perpendicular to the axis and moving in the longitudinal direction synchronously with the rotation of the helices.

Our proposed and implemented prototype helical hybrid microundulator is assembled from four helices manufactured from solid pieces of magnetic materials. The NdFeB helices were manufactured with high accuracy using Wire Electric Discharge machining and magnetized in the axial direction in a strong field of a pulsed solenoid. This reduces the field errors inherent in conventional multi-block Halbach gratings, ensuring the reliability and simplicity of the design. Such a combination of high field strength, small period, as well as manufacturing and assembling simplicity may be effective for implementation of ultra-compact, high-brightness X-ray FELs. Genesis \cite{ref23,ref24}-based calculations using the electron bunch parameters considered in \cite{ref8} show that the use of a hybrid helical microundulator can provide very high peak power and efficiency of the ultra-compact XFEL at a radiation wavelength of 5 Å.

\section*{Acknowledgments}

This research was supported by the Ministry of Innovation, Science and Technology of Israel. The authors thank Moshe Klein for his valuable technical assistance in developing the experimental demonstration system. We are also grateful to E. Bamberg and N. Young for their assistance in fabricating the NdFeB helices for the microundulators, and to A. Weinberg for his help with simulations based on the Genesis code.


\begin{thebibliography}{99}

\bibitem{ref1} P. Emma, R. Akre, J. Arthur, et al., First lasing and operation of an ångstrom wavelength free-electron laser, Nat. Photonics 4 (2010) 641--647.

\bibitem{ref2} T. Ishikawa, H. Aoyagi, T. Asaka, et al., A compact X-ray free-electron laser emitting in the sub-ångström region, Nat. Photonics 6 (2012) 540--544.

\bibitem{ref3} H.-S. Kang, C.K. Min, H. Heo, et al., Hard X-ray free-electron laser with femtosecond-scale timing jitter, Nat. Photonics 11 (2017) 708--713.

\bibitem{ref4} C.J. Milne et.al., SwissFEL: The Swiss X-ray Free Electron Laser, Appl. Sci. 7 (2017) 720.

\bibitem{ref5} B. Faatz et al., The FLASH Facility: Advanced Options for FLASH2 and Future Perspectives, Appl. Sci. 7 (2017) 1114.

\bibitem{ref6} T. Tschentscher et al., Photon Beam Transport and Scientific Instruments at the European XFEL, Appl. Sci. 7 (2017) 592.

\bibitem{ref7} K. Halbach, Physical and optical properties of rare earth cobalt magnets, Nucl. Instrum. Methods Phys. Res. 187 (1981) 109--117.

\bibitem{ref8} J. B. Rosenzweig, N. Majernik, R. R. Robles et al., An ultra-compact x-ray free-electron laser, New J. Phys. 22 (2020) 093067.

\bibitem{ref9} N. Balal, V. L. Bratman, Yu. Lurie, and E. Magory, Efficiency of terahertz undulator radiation from short electron bunches moving in the field of permanently magnetized helices, Phys. Plasmas 28 (2021) 093301.

\bibitem{ref10} E. Magory, N. Balal, and V. L. Bratman, Undulators consisting of magnetized helices, IEEE Trans. Electron Devices 69 (2022) 798.

\bibitem{ref11} N. Balal, E. Bamberg, V. L. Bratman, and E. Magory, Fabrication and Experimental Study of Prototype NdFeB Helical Undulators, IEEE Trans. Electron Devices 70 (2023) 5911.

\bibitem{ref12} E. Magory, V. L. Bratman, and N. Balal, Permanent helical microundulators for x-ray free electron lasers, Phys. Plasmas 31 (2024) 053301.

\bibitem{ref13} F. H. O'Shea, G. Marcus, J.B. Rosenzweig et al., Short period, high field cryogenic undulator for extreme performance x-ray free electron lasers, Phys. Rev. Spec. Top. Accel Beams 13 (2010) 070702.

\bibitem{ref14} Y. Ivanyushenkov, K. Harkay, M. Borland, et al., Development and operating experience of a 1.1-m-long superconducting undulator at the Advanced Photon Source, Phys. Rev. Accel. Beams 20 (2017) 100701.

\bibitem{ref15} T. Tanaka and A. Kagamihata, Demonstration of high performance pole pieces made of monocrystalline dysprosium for short-period undulators, J. Synchrotron Radiat. 26 (2019) 1220--1225.

\bibitem{ref16} S. Casalbuoni, S. Abeghyan, L. Alanakyan, et al., Superconducting undulator activities at the European X-ray free-electron laser facility, Front. Phys. 11 (2023) 1204073.

\bibitem{ref17} J. C. Huang, H. Kitamura, C. S. Yang, Performance investigation of conduction-cooled cryogenic permanent magnet undulator at high beam currents, Phys. Rev. Accel. Beams 27 (2024) 023501.

\bibitem{ref18} A. F. Habib, G. G. Manahan, P. Scherkl et al., Attosecond Angstrom free-electron-laser towards the cold beam limit, Nat. Commun. 14 (2023) 1054.

\bibitem{ref19} N. Majernik and J. Rosenzweig, Design of comb fabricated Halbach undulators, Instruments 3 (2019) 58.

\bibitem{ref20} Cheng-Ying Kuo, Cheng-Hsing Chang, Ting-Yi Chung et al., Design of a Short-Period Helical Permanent Magnet Undulator, IEEE Trans. Appl. Supercond. 28 (2018) 4100805.

\bibitem{ref21} CST Studio Suite®, Version 2025, CST -- Computer Simulation Technology GmbH, Darmstadt, Germany.

\bibitem{ref22} N. Balal, I. V. Bandurkin, V. L. Bratman, and A. E. Fedotov, Helical undulator based on partial redistribution of uniform magnetic field, Phys. Rev. Accel. Beams 20 (2017) 122401.

\bibitem{ref23} S. Reiche, Genesis 1.3: A fully 3D time-dependent FEL simulation code, (1999).

\bibitem{ref24} S. Reiche and C. Lechner, Status of the time-dependent FEL code Genesis 1.3 (Version 4), Proc. IPAC'24, WEPR58, Nashville, 2024.

\bibitem{ref25} V.L. Bratman, N.S. Ginzburg, M.I. Petelin, Nonlinear theory of stimulated wave scattering by relativistic electron beams, Sov. Phys. JETP 49 (1979) 469--475.

\bibitem{ref26} R. Bonifacio, C. Pellegrini and I. Narducci, Collective instabilities and high-gain regime in a free electron laser, Opt. Commun. 30 (1984) 373--378.

\bibitem{ref27} A. Weinberg and A. Nause, Dogleg design for an MeV ultra-fast electron diffraction beam-line for the hybrid photo-emitted RF gun at Ariel University, Nucl. Instrum. Methods Phys. Res., Sect. A 989 (2021) 164952.

\end{thebibliography}
\end{document}